# Envy-Free Makespan Approximation

Edith Cohen[*]    Michal Feldman[†]    Amos Fiat[‡]    Haim Kaplan [§]

Svetlana Olonetsky[¶]

*"It is better to be envied than pitied"*

— Herodotus (484 BC - 430 BC)

**Abstract**

We study envy-free mechanisms for scheduling tasks on unrelated machines (agents) that approximately minimize the makespan. For indivisible tasks, we put forward an envy-free poly-time mechanism that approximates the minimal makespan to within a factor of $O(\log m)$, where $m$ is the number of machines. We also show a lower bound of $\Omega(\log m / \log \log m)$. This improves the recent result of Hartline *et al.* [15] who give an upper bound of $(m+1)/2$, and a lower bound of $2-1/m$. For divisible tasks, we show that there always exists an envy-free poly-time mechanism with optimal makespan.

## 1 Introduction

Imagine a set of $n$ household chores, and $m$ kids in the family. Every chore may take a different amount of time to be performed by each child. A single chore cannot be performed by more than one child (indivisible), but multiple chores can be assigned to a single child. The parents want to allocate chores *fairly*, and may offer inducements to the children so as to ensure fairness. The parents have an additional goal which is to get all the chores out of the way as soon as possible. This problem is our main focus. In job scheduling terminology, we study mechanisms for the fair allocation of tasks to machines (agents), each of which may take a different length of time to complete every task, subject to the added goal of minimizing the makespan; *i.e.*, getting all tasks done as soon as possible.

The problem of fair division, often modeled as that of partitioning a cake fairly, goes back at least to 1947 and is attributed to Jerzy Neyman, Hugo Steinhaus, Stefan Banach and Bronislaw Knaster ([24, 25]). There are several books on fair division, and hundreds of references, both mathematical and philosophical, a small sampling of books is [26, 4, 20, 16, 23]. Martin Gardner (1978, [12]) is credited with asking about fair division of household chores.

In order to devise a fair division, one should first define the precise notion of fairness desired. One common notion of fairness is that of "envy-freeness", an allocation where no one seeks to switch her outcome with that of another (Dubins and Spanier, 1961, [10], Foley, 1967, [11]).

---

[*]AT&T Labs-Research, 180 Park Avenue, Florham Park, NJ.

[†]School of Business Administration, The Hebrew University of Jerusalem.

[‡]The Blavatnik School of Computer Science, Tel Aviv University.

[§]The Blavatnik School of Computer Science, Tel Aviv University.

[¶]The Blavatnik School of Computer Science, Tel Aviv University.

Tasks may be divisible or indivisible. It is always possible to divide a divisible task equally between all agents. This is envy-free, but infinite task times (e.g., a task too demanding for a four year old) may make this assignment impossible or ill-defined.

For indivisible tasks, it is less obvious that one can achieve envy-freeness. This can be achieved if one allows for the design of a *mechanism* that determines both an allocation and payments (to or from the agents, to the mechanism or between themselves). We assume that an agent's utility is quasi-linear, *i.e.*, equal to the payment from the mechanism from which we subtract the cost of tasks assigned by the mechanism. In particular, assigning task $j$ to the agent requiring minimal time for $j$ (maximizing social welfare) and using VCG payments makes this task assignment envy-free.

Within the range of possible envy-free allocations, one may seek out an envy-free allocation that achieves additional goals, such as economic efficiency, revenue maximization and incentive compatibility.[1]

Hartline *et al.* studied the additional goal of makespan minimization. In particular, they seek envy-free mechanisms for scheduling (indivisible) tasks on unrelated machines (agents) that approximately minimize the makespan. Consider an instance of indivisible task scheduling for $m$ agents (without envy-free requirements), and without loss of generality assume that the assignment of minimal makespan has makespan 1. Hartline *et al.* show that there is no envy-free mechanism that guarantees a makespan less than $2 - 1/m$. They also give an algorithm that always produces a schedule with makespan at most $(m+1)/2$.

Hartline *et al.* also define *locally efficient* allocations, and prove that this is a necessary and sufficient condition that such an allocation has associated payments that make it envy-free. (The locally efficient condition is more general than the context of task scheduling).

The above setting of job scheduling on unrelated machines has been first proposed by Nisan and Ronen [22] in their seminal paper on *incentive compatible* mechanisms. Nisan and Ronen were not concerned with the fairness of the allocation, rather they looked for an incentive compatible mechanism that approximates the minimal makespan. The problem posed by Nisan and Ronen has led to a great deal of work [18, 8, 17, 1], yet the main question is still open. For the general case, all that known is a lower bound of 2.61 and an upper bound of $m$ (similar to the gap obtained by Hartline *et al.* for envy-free mechanisms)[2]. For divisible tasks, Christodoulou *et al.* [7] demonstrated an upper bound of $1 + (m-1)/2$ and a lower bound of $2 - 1/m$ (while for the class of "task independent" algorithms, the bound of $1 + (m-1)/2$ holds as a lower bound as well).

## 1.1 Our Contributions

- We give an envy-free mechanism for scheduling indivisible tasks on $m$ unrelated machines, that approximates the minimal makespan to within a factor of $O(\log m)$, improving the $(m+1)/2$ of [15]. Our mechanism is polynomial time. (Section 3)

- We give a lower bound of $\Omega(\log m / \log \log m)$ on the makespan approximation of any envy-free indivisible tasks scheduling mechanism, polynomial time, or not. This improves on the previous $2 - 1/m$ of [15]. (Section 4)

---
[1] Several papers consider envy-free item pricing (in various scenarios) with the goal of maximizing revenue [14, 6, 5, 2], hardness results for revenue maximization envy-free item pricing appear in [9].

[2]The bounds given above hold for deterministic mechanism, but randomization can reduce the approximation ratio. In particular, Mualem and Schapira [21] advocated a randomized truthful mechanism with an upper bound of $7m/8$ and showed a lower bound of $2 - 1/m$ for randomized mechanisms.

- For machine scheduling with *divisible* tasks, we show that there always exists an envy-free polynomial-time mechanism with optimal makespan (Section 5).

## 2 Preliminaries

In the scheduling notation of [13], the input to the problem $(R||C_{max})$ is defined as follows: there are $m$ machines, $n$ tasks, and a matrix $(c_{ij})_{1 \leq i \leq m, 1 \leq j \leq n}$ such that $c_{ij}$ is the time (load or cost) of running task $j$ on machine $i$.

Machine scheduling can have either divisible or indivisible tasks. An assignment of tasks to machines is specified by an $m \times n$ matrix $a = (a_{ij})$, where $a_{ij}$ is the fraction of job $j$ assigned to machine $i$. A valid assignment must have $\sum_{j \in [m]} a_{ij} = 1$ for all jobs $i \in [n]$. If tasks are divisible then $0 \leq a_{ij} \leq 1$, for indivisible tasks $a_{ij} \in \{0, 1\}$.

Let $\bar{c}_i = (c_{i1}, \ldots, c_{in})$ be the $i$'th row of the cost matrix $c = (c_{ij})$ and let $\bar{a}_i$ be the $i$'th row of the assignment matrix $a = (a_{ij})$. The load (cost) of machine $i$ under assignment $(a_{ij})$ is the inner product $\bar{c}_i \cdot \bar{a}_i = \sum_{j=1}^n c_{ij} a_{ij}$. The makespan of the assignment matrix $a$ is $\max_{1 \leq i \leq m} \bar{c}_i \cdot \bar{a}_i$.

The problem of finding an assignment with a minimum makespan can be formulated as an integer program (IP) for indivisible jobs ($a_{ij} \in \{0, 1\}$) and as a linear program (LP) for divisible jobs ($0 \leq a_{ij} \leq 1$). Lenstra, Shmoys and Tardos ([19]) give a polynomial time 2-approximation algorithm for scheduling indivisible tasks, and an inapproximability result, stating that unless $P = NP$, for $\rho < 3/2$ there is no polynomial time $\rho$-approximation algorithm for minimizing makespan of indivisible tasks.

Following [22, 15], we consider the setting where the $m$ machines are selfish agents. An allocation function $a$ maps the cost matrix $c = (c_{ij})$ into an $m \times n$ assignment matrix $a(c) = (a_{ij})$. Let $\bar{c}_i$ and $\bar{a}(c)_i$ be the $i$'th row of row the matrices $c$ and $a(c)$, respectively. A payment function $p$ is a mapping from the $m \times n$ cost matrix $c$ to a real vector $p(c) = (p_1, p_2, \ldots, p_m)$, $p_i \in \Re$. Let $p(c)_i$ be the $i$'th coordinate of $p(c)$.

A mechanism is a pair of functions, $M = <a, p>$, where $a$ is an allocation function, and $p$ is a payment function. For mechanism $<a, p>$ with cost matrix $c = (c_{ij})$, the utility to agent $i$ is $p(c)_i - \bar{c}_i \cdot \bar{a}(c)_i$. Such a utility function is known as quasi-linear.

A mechanism $<a, p>$ is *envy-free* if no agent seeks to switch her allocation and payment with another. I.e., if for all $1 \leq i, j \leq m$ and all cost matrices $c$:

$$p(c)_i - \bar{c}_i \cdot \bar{a}(c)_i \geq p(c)_j - \bar{c}_i \cdot \bar{a}(c)_j.$$

Based on [15], we say that an allocation function $a$ is *envy-free implementable* (*EF*-implementable) if there exists a payment function $p$ such that the mechanism $M = <a, p>$ is envy-free.

**Characterization**

We make use of the following definition and theorem from Hartline *et al.* [15]:

An allocation function $a$ is said to be *locally efficient* if for all cost matrices $c$ and all permutations $\pi$ of $1, \ldots, m$,

$$\sum_{i=1}^m \bar{c}_i \cdot \bar{a}(c)_i \leq \sum_{i=1}^m \bar{c}_i \cdot \bar{a}(c)_{\pi(i)}.$$

**Theorem 2.1.** *([15]) A necessary and sufficient condition for an allocation function $a$ to be EF-implementable is that assignment $a$ is locally efficient.*

# 3  An Upper Bound for Indivisible Jobs

In this section we present a polynomial algorithm that produces a locally efficient, and hence, envy-free allocation of indivisible jobs whose makespan is at most $O(\log m)$ times the optimal makespan without envy-freeness constraints. In particular, our algorithm approximates the minimum makespan with envy-freeness constraints to within a factor of $O(\log m)$.

To simplify the description we assume that the algorithm starts with an allocation OPT that minimizes the makespan. If we were to start with an $\alpha$ approximation to the minimal makespan, ([19]), the final approximation would be $2\alpha \cdot e(\ln m + 1) = O(\log m)$. The allocation, which we start with, fixes a partition of the jobs into bundles[3] $B = \{b_1, \ldots, b_m\}$ where $b_i$ is the set of jobs running on machine $i$. In addition to set notation ($a_i$ is a set of tasks) we use vector notation ($\bar{a}_i$ is a 0/1 vector of length $n$, the $j$'th coordinate of which is one iff task $j$ belongs to $a_i$).

We use OPT to denote both the allocation and its makespan when no confusion can arise. For set of bundles $D = \{d_j\}_{j=1}^k$, $k \leq m$, we denote by $LE(D)$ a locally efficient assignment of $D$ (this is an assignment of bundles to machines, no more than one bundle per machine, such that the sum of the loads is minimized).

The algorithm works in phases. We start each phase with a subset of the bundles that have not been assigned to machines yet. We compute a locally efficient assignment of these bundles. Then if this locally efficient assignment contains a machine with load larger than 2OPT we discard all bundles assigned to such machines (these bundles will be considered again only in the next iteration), and repeat the process with the remaining bundles until the makespan of the locally efficient allocation is at most 2OPT. Thus, each phase produces an assignment of some subset of the bundles. The final assignment is the union of the assignments obtained in the different phases. Specifically, we assign to each machine the union of the bundles assigned to it in the different phases. See Algorithm FIND-APPROX.

We now prove the following theorem.

**Theorem 3.1.** *The allocation computed in Algorithm* FIND-APPROX *is locally efficient and its makespan is $O(\log m \cdot OPT)$.*

The following lemma shows that in each phase the number of bundles which we discard is at most half the number of bundles we start out with.

**Lemma 3.2.** *During a phase of Algorithm* FIND-APPROX *(lines 5-20) that starts with $k$ bundles, no more than $k/2$ bundles are discarded.*

*Proof.* Consider the set of bundles $B_{active} = \{b_{i_1}, \cdots, b_{i_k}\}$, $k = |B_{active}|$, at the beginning of a phase. Let $a_i$ be the bundle assigned to machine $i$ by the locally efficient assignment $LE(B_{active})$. It follows that

$$\sum_{i=1}^m \bar{c}_i \cdot \bar{a}_i \leq \sum_{j=1}^k \bar{c}_{i_j} \cdot \bar{b}_{i_j} \leq k \cdot \text{OPT} .$$

Every time we throw out bundles in the inner loop (lines 8-16 of Algorithm FIND-APPROX ) and recompute the assignment of the remaining bundles $\sum_i \bar{c}_i \cdot \bar{a}_i$ decreases by at least $2 \cdot \text{OPT}$. Since $\sum_i \bar{c}_i \cdot \bar{a}_i$ never increases during a phase, the inner loop can repeat at most $\frac{k \cdot \text{OPT}}{2 \cdot \text{OPT}} = \frac{k}{2}$ times, implying that at most $\frac{k}{2}$ bundles can join the set $B_{out}$.  □

---
[3] In this paper, bundles consist of jobs or other objects, and do not include payments which are dealt with separately.

**Algorithm 1** Compute Envy-Free ($O(\log m)$)-Approximation

```
 1: procedure FIND-APPROX(B, c)                    ▷ B – set of OPT bundles, c – cost matrix
 2:     q ← 0
 3:     B_out ← ∅
 4:     B_active ← B
 5:     while B_active ≠ ∅ do
 6:         q ← q + 1
 7:         a ← LE(B_active)
 8:         while makespan(a) > 2 · OPT do
 9:             for all i do
10:                 if c̄_i · ā_i > 2 · OPT then
11:                     B_out ← B_out ∪ a_i
12:                     B_active ← B_active \ a_i
13:                 end if
14:             end for
15:             a ← LE(B_active)
16:         end while
17:         a^q ← a
18:         B_active ← B_out
19:         B_out ← ∅
20:     end while
21:     a_i = ∪_{j=1}^q a_i^j
22:     return a
23: end procedure
```

The following lemma follows directly from Lemma 3.2.

**Lemma 3.3.** *When Algorithm* FIND-APPROX *terminates* $q \leq \log m + 1$.

It follows from the definition of the algorithm that the makespan of the assignment $a^j$ produced by phase $j$ is at most 2OPT. The final assignment assigns to each machine the union of the bundles assigned to it by the different phases. Since we have at most $\log m + 1$ phases we obtain that the makespan of the final assignment is $O(\log m \cdot \text{OPT})$. To finish the proof of Theorem 3.1 we have to show that the assignment which we produce is locally efficient. This follows from a more general observation that any union of locally efficient assignments is locally efficient as established by the following lemma.

**Lemma 3.4.** *Let $c$ be a cost matrix, and let $b$ and $b'$ be two assignments of different sets of jobs to the same set of machines. Let $a$ be the assignment such that for every $i$, $a_i = b_i \cup b'_i$. If $b$ and $b'$ are locally efficient then so is $a$.*

*Proof.* Assume that $a$ is not locally efficient then there is a permutation $\pi$ of $1, 2, \ldots, m$ such that $\sum \bar{c}_i \cdot \bar{a}_{\pi(i)} < \sum \bar{c}_i \cdot \bar{a}_i$. By the definition of $a$, this implies that $\sum (\bar{c}_i \cdot \bar{b}_{\pi(i)} + \bar{c}_i \cdot \bar{b}'_{\pi(i)}) < \sum (\bar{c}_i \cdot \bar{b}_i + \bar{c}_i \cdot \bar{b}'_i)$ and, therefore, either $\sum \bar{c}_i \cdot \bar{b}_{\pi(i)} < \sum \bar{c}_i \cdot \bar{b}_i$ or $\sum \bar{c}_i \cdot \bar{b}'_{\pi(i)} < \sum \bar{c}_i \cdot \bar{b}'_i$, which either contradicts the assumption that $b$ is locally efficient or contradicts the assumption that $b'$ is locally efficient. □

**Remark1:** We can replace the constant 2 in lines 8 and 10 of Algorithm FIND-APPROX by the constant $e$. Then the number of iterations is at most $\ln m$ and the makespan would be at most $e(\ln m + 1)$. Note that $e \ln m < 2 \log_2 m$.

**Remark2:** In order to get polynomial running time for Algorithm FIND-APPROX we can start with any constant approximation to makespan. Locally efficient assignment given bundles can be calculated using weighted matching in polynomial time.

## 4 A Lower Bound for Indivisible Jobs

We give a lower bound of $\Omega(\log m / \log \log m)$ on the makespan approximation achievable by any locally efficient allocation of indivisible jobs.

Let $n$ be the number of jobs. For every $n$ we define the cost matrix $c = (c_{ij})$ with $m = n + \ell$ machines where $2^\ell = \log n / (4 \log \log n)$ as follows.

$$c = \begin{pmatrix} 1 & \infty & \infty & \infty & \cdots & \infty & \infty \\ 1 - \frac{1}{2(n-1)} & 1 & \infty & \infty & \cdots & \infty & \infty \\ 1 - \frac{2}{2(n-1)} & 1 - \frac{1}{2(n-2)} & 1 & \infty & \cdots & \infty & \infty \\ 1 - \frac{3}{2(n-1)} & 1 - \frac{2}{2(n-2)} & 1 - \frac{1}{2(n-3)} & 1 & \cdots & \infty & \infty \\ \vdots & \vdots & \vdots & & & & \vdots \\ 1/2 + \frac{1}{2(n-1)} & 1/2 + \frac{1}{2(n-2)} & 1/2 + \frac{1}{2(n-3)} & 1/2 + \frac{1}{2(n-4)} & \cdots & 1 & \infty \\ 1/2 & 1/2 & 1/2 & 1/2 & \cdots & 1/2 & 1 \\ \hline 2 & 2 & 2 & 2 & \cdots & 2 & 2 \\ 4 & 4 & 4 & 4 & \cdots & 4 & 4 \\ \vdots & \vdots & \vdots & & & \vdots & \vdots \\ 2^\ell & 2^\ell & 2^\ell & 2^\ell & \cdots & 2^\ell & 2^\ell \end{pmatrix}$$

Row $i$, $1 \leq i \leq n + \ell$, of the cost matrix $c$ corresponds to the $i$th machine and $c_{ij}$ is the cost of running job $j$ on machine $i$. The horizontal line lies between machine $n$ and $n+1$. For $1 \leq i \leq n$, machine $i$ has cost 1 for job $i$, costs $1 - (i-j)/(2(n-j))$ for jobs $j < i$, and cost of $\infty$ for the rest of jobs ($j > i$). For $n + 1 \leq i \leq n + \ell$, all costs of machine $i$ are equal to $2^i$. Observe that $c_{ij} - c_{i+1,j} = 1/(2(n-j))$ for $1 \leq i < n$ and $j \leq i$.

The optimal makespan of all these matrices is 1. We can achieve makespan 1 by running job $i$ on machine $i$ for every $1 \leq i \leq n$, and we cannot do better since job $n$ has load $\geq 1$ on all machines.

We establish a lower bound of $2^\ell = \log n / (4 \log \log n)$ on the makespan of any envy-free allocation for this instance. This shows that we cannot have an algorithm that always produces an envy-free allocation whose makespan approximates the optimal makespan to within a factor smaller than $\log n / (4 \log \log n)$.

Specifically, we show that for *any* partition of the jobs into $\leq n + \ell$ bundles, any locally efficient assignment of these bundles to the machines has makespan at least $2^\ell$. Our first lemma makes few easy observations regarding allocations with makespan $< 2^\ell$.

**Lemma 4.1.** *For cost matrix $(c_{ij})$ above, any allocation with makespan $< 2^\ell$ satisfies the following.*

1. *There are fewer than $2^{\ell+1}$ jobs on each machine.*

2. *There are fewer than $2^\ell / 2^{(i-n)}$ jobs on machine $n < i \leq n + \ell$.*

3. There are fewer than $2^\ell$ jobs running on machines $n+1, \ldots, n+\ell$.

*Proof.* (1) follows since $c_{ij} \geq 1/2$ for all $i$ and $j$. (2) follows since for $n < i \leq n+\ell$, $c_{ij} \geq 2^{i-n}$. (3) follows by summing the upper bound on the number of jobs on each of these machines, this sum is $\leq \sum_{i=n+1}^{(n+\ell)} (2^\ell/2^{(i-n)} - 1) = 2^\ell - \ell < 2^\ell$. □

We can now conclude with the proof of the lower bound:

**Theorem 4.2.** *For any partition of jobs into bundles, the makespan of any locally efficient assignment of the bundles is at least $2^\ell = \log n/(4 \log \log n)$.*

*Proof.* Fix an arbitrary partition into bundles and suppose that there is an envy-free assignment of the bundles with makespan $< 2^\ell$. We will derive a contradiction by showing that the assignment is not locally efficient.

Since the makespan is $< 2^\ell$ no bundle is assigned to machine $n+\ell$. So we derive the contradiction by showing that if we move the bundle assigned to machine $i$ to machine $i+1$ for $1 \leq i < n+\ell$ we decrease the total load.

By Lemma 4.1(1), there are $\leq 2^{\ell+1} - 1$ jobs in the bundle assigned to machine $n$. So moving this bundle to machine $n+1$ increases the load of this bundle by at most $3/2 \cdot 2^{\ell+1}$.

Since $c_{i+1,j} = 2c_{ij}$ for $n+1 \leq i < n+\ell$ and all $j$, moving a bundle from machine $i$ to machine $i+1$ for $n+1 \leq i < n+\ell$ increases the load of this bundle exactly by a factor of 2. Since the load on each of these machines is $< 2^\ell$, the total increase in load caused by moving each of these bundles one machine down is $< \ell \cdot 2^\ell$.

Summing up we obtain that the increase in the load due to moving bundles on machines $n, \ldots, n+\ell$ is at most $(3/2)2^{\ell+1} + \ell 2^\ell = (\ell+3)2^\ell$. Substituting $2^\ell = \log n/(4 \log \log n)$ we obtain that this increase for large $n$ is smaller than $\log n/4$ which is smaller than, say, $\ln n/2.5$.

By Lemma 4.1(3), at most $2^\ell$ jobs are in bundles assigned to machine $n+1, \ldots, n+\ell$ and, by Lemma 4.1(1), at most $2^{\ell+1}$ jobs are in the bundle assigned to machine $n$. Therefore, at least $n - 3 \cdot 2^\ell$ jobs are in bundles assigned to machines $1, \ldots, n-1$. If job $j$ is in one of these bundles then after we move these bundles the contribution of job $j$ to the load decreases by $1/(2(n-j))$. So the total decrease in load due to moving bundles assigned to machines $1, \ldots, n-1$ is at least $(1/2)(H_n - H_{3 \cdot 2^\ell}) \approx (1/2)(\ln n - \ln(3 \cdot 2^\ell)) \geq (1/2 - \epsilon) \ln n$ for large enough $n$.

Since the decrease in the load caused by moving bundles on machines $1, \ldots, n-1$ is larger than the increase in the load caused by moving bundles on machines $n, \ldots, n+\ell$ we obtain a contradiction. □

Since $m = n + \ell = O(n)$, we get that it is $\Omega(\log m/\log \log m)$ approximation.

## 5 Unrelated Machine Scheduling with Divisible Jobs

The existence of an envy-free assignment for divisible tasks is trivial, even without payments. For example, simply assigning each machine a $1/m$ fraction of every job is trivially envy-free. However, it is certainly not optimal with respect to makespan minimization. In the previous section we showed a lower bound of $\Omega(\log m/\log \log m)$ for indivisible tasks.

In this section we prove that when tasks are divisible, there always exists an envy-free allocation that achieves the minimum makespan. To find such an allocation: solve the linear program that minimizes makespan subject to the constraints of a valid assignment including envy-free constraints. The main issue is to prove that this LP formulation has a solution.

**Theorem 5.1.** *For any instance of machine scheduling with divisible jobs, there is a locally efficient assignment with minimum makespan.*

Consider an instance of the machine scheduling problem, specified by the cost matrix $c = (c_{ij})$. We use the notation $\bar{c}_i$ for the $i$'th row of the cost matrix. As we deal with indivisible assignments, bundles are sets of fractions of tasks. An assignment itself is represented as a real valued matrix, $(a_{ij})$, where $a_{ij}$ is the fraction of task $j$ assigned to machine $i$. We use the terminology of sets ($a_i$ is the set of fractional tasks assigned to agent $i$) as well as vector notation ($\bar{a}_i$ is the $i$'th row of this assignment matrix $(a_{ij})$).

**Warm up (two machines with finite valuations):** Consider an instance with two machines $i \in [2]$ such that all entries in $c$ are finite. We show that *any* assignment with minimum makespan must be locally efficient. Let $o$ be an optimal assignment and assume on the contrary that $o$ is not locally efficient. Without loss of generality, we can assume that the makespan of $o$ is 1 and both machines have the same load $\bar{c}_1 \cdot \bar{o}_1 = \bar{c}_2 \cdot \bar{o}_2 = 1$, where $o_i$ is the bundle assigned to machine $i$ ($i \in [2]$). (Otherwise, we can transfer (fractional) jobs from the machine with load 1 to the other machine and get an assignment $a$ with a strictly lower makespan than $o$, which contradicts optimality of $o$.)

Consider a locally efficient solution $e$ with bundles $o_1$ and $o_2$. Since $o \neq e$, $e_1 = o_2$ and $e_2 = o_1$. The sum of the loads under $e$ must be strictly smaller than 2, which is the sum of the loads under $o$ (this is because $o$ is not locally efficient). The makespan of $e$ must be at least 1 (otherwise, $e$ has smaller makespan than $o$ which contradicts optimality). Therefore, under $e$, exactly one of the machines must have load strictly smaller than 1. Without loss of generality, we assume it is machine 1 and let $\bar{c}_2 \cdot \bar{e}_2 = \bar{c}_2 \cdot \bar{o}_1 = 1 + x$ and $\bar{c}_1 \cdot \bar{e}_1 = \bar{c}_1 \cdot \bar{o}_2 = 1 - y$, where $x \geq 0$ and $0 \leq y \leq 1$. The sum of the loads under $e$ is $1 + x + 1 - y < 2$. Hence, $y > x$.

We now construct a new assignment $a$ such that $\bar{a}_1 = \bar{o}_2 + (y - \epsilon)\bar{o}_1$ and $\bar{a}_2 = (1 - y + \epsilon)\bar{o}_1$. It is easy to see that $a$ is well defined for any $0 < \epsilon < y$. We show that $a$ has makespan $\max\{\bar{c}_1 \cdot \bar{a}_1, \bar{c}_2 \cdot \bar{a}_2\} < 1$, which contradicts optimality of $o$. Using $\epsilon = (y-x)/2$, the load of $a$ on machine 1 is $\bar{c}_1 \cdot \bar{a}_1 = (1-y) + (y-\epsilon) = 1 - \epsilon < 1$ and on machine 2 is $\bar{c}_2 \cdot \bar{a}_2 = (1 - y + \epsilon)(1+x) = 1 - y + x - yx + \epsilon + \epsilon x \leq 1 - (y-x) + \epsilon(1+x) < 1 - (y-x) + 2\epsilon = 1$.

**General instance:** Consider a cost matrix $c = (c_{ij})$ of the machine scheduling problem with $m \geq 2$ machines which may include $+\infty$ entries. We first define a lexicographic order on assignments.

**Definition 5.2.** *A vector $(l_1, \ldots, l_m)$ is smaller than $(l'_1, \ldots, l'_m)$ lexicographically if for some $i$, $l_i < l'_i$ and $l_k = l'_k$ for all $k < i$. An assignment $a$ is smaller than $a'$ lexicographically if the vector of machine loads $l(a) = (l_1(a), \ldots, l_m(a))$, sorted in non-increasing order, is lexicographically smaller than $l(a')$, sorted in non increasing order.*

Clearly, every lexicographically minimal assignment has minimum makespan. When all entries are finite, any assignment with minimum makespan has equal loads on all machines and therefore minimum makespan implies a lexicographically minimal assignment.[4] In either case (with all entries finite or not), there exists some lexicographically minimal schedule with minimal makespan. In order to prove Theorem 5.1, it suffices for us to prove that a lexicographically minimal schedule is also locally efficient.

---

[4]To see this, suppose on the contrary that this is not the case. Consider an assignment with minimum makespan. Let $\mathcal{M}' \subset [m]$ be machines with load strictly lower than the makespan. We can transfer (fractional) jobs from $[m] \setminus \mathcal{M}'$ machines to $\mathcal{M}'$ machines and obtain an assignment with strictly lower makespan, which contradicts optimality.

**Lemma 5.3.** *Every lexicographically minimal assignment is locally efficient.*

*Proof.* Assume on the contrary that $o$ is a lexicographically minimal assignment that is not locally efficient and let $e$ be a locally efficient assignment of the bundles of $o$. Consider a directed graph $G$ between machines where arcs correspond to a reassignment of bundles between $o$ and $e$. We also include empty bundles and hence this reassignment is a permutation. Each machine has either no incoming and outgoing arcs or exactly one incoming and exactly one outgoing arc. The graph $G$ is therefore a collection of singletons and cycles.

Since $e \neq o$, and there are no paths, $G$ must contain a cycle. Moreover, because $e$ is locally efficient and $o$ is not, $G$ must contain a cycle $X$ such that

$$\sum_{i \in X} \bar{c}_i \cdot \bar{o}_i > \sum_{i \in X} \bar{c}_i \cdot \bar{e}_i \ . \tag{1}$$

Consider such a cycle $X$ with $|X| = k \geq 2$ nodes (machines). We (re-)number machines such that machines along the cycle are indexed $[0, \ldots, k-1]$ in cyclic order. We also accordingly renumber bundles such that the bundles of machine $i$ are $o_i, e_i$. By definition, $e_{i+1} = o_i$ (all addition operations through this section are modulo $k$).

We claim that all machines on the cycle $X$ must be equally loaded under $o$, that is, $\bar{c}_i \cdot \bar{o}_i$ are equal for $0 \leq i \leq k-1$.[5] Assume on the contrary that there are two consecutive machines on $X$, $i$ and $i+1$, such that $\bar{c}_i \cdot \bar{o}_i > \bar{c}_{i+1} \cdot \bar{o}_{i+1}$. We construct an assignment $a$ from $o$ by shifting a fraction $f$ of the bundle $o_i$ from machine $i$ to machine $i+1$ such that both machines have equal loads, that is, $f$ such that $(1-f)\bar{c}_i \cdot \bar{o}_i = \bar{c}_{i+1} \cdot (\bar{o}_{i+1} + f\bar{o}_i)$ or explicitly $f = \frac{\bar{c}_i \cdot \bar{o}_i - \bar{c}_{i+1} \cdot \bar{o}_{i+1}}{(\bar{c}_i + \bar{c}_{i+1}) \cdot \bar{o}_i}$. Since $e_{i+1} = o_i$ and $\bar{c}_{i+1} \cdot \bar{e}_{i+1} < \infty$, $\bar{c}_{i+1} \cdot \bar{o}_i < \infty$ and therefore $f > 0$. The assignment $a$ is strictly lexicographically smaller than $o$, which is a contradiction.

By scaling, we can assume that $\bar{c}_i \cdot \bar{o}_i = 1$ for all $0 \leq i \leq k-1$. Let $\bar{c}_{i+1} \cdot \bar{e}_{i+1} = 1 + \Delta_i$ be the load of machine $i+1$ under $e$ ($\Delta_i \geq -1$.) From (1),

$$\sum_{i=0}^{k-1} \Delta_i < 0 \ . \tag{2}$$

We conclude the proof by constructing an assignment $a$ that is identical to $o$ outside the cycle, has $\sum_{i=0}^{k-1} \bar{a}_i = \sum_{i=0}^{k-1} \bar{o}_i$, that is, same total allocation as $o$ on cycle machines, has load $\bar{c}_i \cdot \bar{a}_i = 1$ on machines $i = 1, \ldots, k-1$ and load $\bar{c}_0 \cdot \bar{a}_0 < 1$ on machine 0. The assignment $a$ is strictly lexicographically smaller than $o$, which contradicts the assumption that $o$ is the lexicographically minimum.

The assignment $a$ is such that for $i = 0, \ldots, k-1$, an $0 \leq \alpha_i \leq 1$ fraction of $o_i$ is assigned to machine $i$ and the remaining $(1 - \alpha_i)$ is assigned to machine $i+1$. Define

$$\mu = \max\{1, \max_{i=0\ldots k-1} \prod_{j=0}^{i}(1 + \Delta_i)\} \ , \tag{3}$$

$\alpha_0 = 1 - \frac{1}{2\mu}$, and for $i = 1, \ldots, k-1$:

$$(1 - \alpha_i) = (1 - \alpha_{i-1})(1 + \Delta_{i-1}) \ . \tag{4}$$

---
[5]This is immediate if there are no $+\infty$ entries in $v$.

It follows that for $i = 1, \ldots, k-1$,

$$(1 - \alpha_i) = (1 - \alpha_0) \prod_{j=0}^{i-1} (1 + \Delta_j) \ . \tag{5}$$

We show that all $\alpha_i$ are well defined (are in $[0,1]$): Since $\infty > \mu \geq 1$, $\alpha_0 \in [1/2, 1)$. For $i = 1, \ldots, k-1$, using (5) and (3)

$$(1 - \alpha_i) = \frac{1}{2\mu} \prod_{j=0}^{i-1} (1 + \Delta_j) \leq \frac{1}{2} < 1 \ . \tag{6}$$

As a product of positive quantities, $(1 - \alpha_i) \geq 0$.

The load $a_i$ takes on machine $i$ ($i = 1, \ldots, k-1$) is (using (4)):

$$\bar{c}_i \cdot \bar{a}_i = \alpha_i + (1 + \Delta_{i-1})(1 - \alpha_{i-1}) = \alpha_i + (1 - \alpha_i) = 1 \ . \tag{7}$$

The load $a_0$ takes on machine 0 is (using (5)):

$$\bar{c}_0 \cdot \bar{a}_0 = \alpha_0 + (1 + \Delta_{k-1})(1 - \alpha_{k-1}) = \alpha_0 + (1 - \alpha_0) \prod_{j=0}^{k-1} (1 + \Delta_j) < \alpha_0 + (1 - \alpha_0) = 1 \ . \tag{8}$$

The strict inequality follows from $(1 - \alpha_0) = 1/(2\mu) > 0$ and from $\prod_{j=0}^{k-1}(1 + \Delta_j) < 1$ (which follows from (2)). □

## 6 Summary and Open Problems

Table 1 summarizes upper and lower bounds on the ratio of the optimal makespan of machine scheduling with envy-freeness constraints and the optimal makespan without envy-freeness constraints. The upper bounds correspond to polynomial time algorithms. An obvious challenge is to close the gap between the upper and lower bounds for indivisible tasks.

|  | Lower bound | Upper bound |
| --- | --- | --- |
| (Divisible+EF)/Divisible | 1 | 1 (Thm. 5.1) |
| (Indivisible+EF)/Indivisible | $\Omega(\frac{\log m}{\log \log m})$ (Thm. 4.2) | $O(\log m)$ (Thm. 3.1) |

Table 1: Summary of our results on the cost of envy-freeness. The rows correspond to divisible or indivisible tasks. The columns correspond to upper bounds on the ratio and lower bounds on the *worst-case* ratio. The number of machines is $m$.

An intriguing issue is to understand the interaction of envy-freeness and incentive compatibility. What can we say about the makespan approximation for mechanisms that are *both* envy-free and incentive compatible? Clearly, any $o(m)$ approximation that is both incentive compatible and envy-free would be a major breakthrough, even without envy-freeness. Recently, Ashlagi, Dobzinski, and Lavi [1] gave a lower bound of $\Omega(m)$ on makespan approximation for incentive compatible and *anonymous* mechanisms. What if we discard the *anonymous* assumption but require that the mechanism also be envy-free?

Minimum makespan machine scheduling is classically formulated as a linear program (for divisible jobs) or an integer program (for indivisible jobs), both with the same set of linear constraints. The requirement of envy-freeness can be captured by adding payment variables (that are not required to be integral) as additional linear constraints. Accordingly, for a cost matrix $(c_{ij})$, we denote the optimal makespan with or without integrality or envy-freeness by $T_{\texttt{LP}}(c_{ij})$, $T_{\texttt{IP}}(c_{ij})$, $T_{\texttt{LP+EF}}(c_{ij})$, and $T_{\texttt{IP+EF}}(c_{ij})$. Using this notation, Table 1 lists bounds on the ratios $T_{\texttt{IP+EF}}(c_{ij})/T_{\texttt{IP}}(c_{ij})$ (indivisible) and $T_{\texttt{LP+EF}}(c_{ij})/T_{\texttt{LP}}(c_{ij})$ (divisible).

Starting with divisible tasks without envy-freeness constraints ($T_{\texttt{LP}}(c_{ij})$) we consider the impact on the optimal makespan of integrality and envy-freeness. Envy-freeness requirement alone does not result in an increase of the optimal makespan (Thm. 5.1). There are instances (the instances in our lower bound construction in Thm. 4.2) where the integrality requirement (indivisible tasks) results in at most a factor 2 increase while, curiously, the combination of *both* requirements results in $\Omega(\log m / \log \log m)$ factor increase.

Considering the approximability of the optimal makespan under the different types of constraints, $T_{\texttt{LP}}$ and $T_{\texttt{LP+EF}}$ are linear programs and hence solvable in polynomial time and $T_{\texttt{IP}}$ has a 2 approximation algorithm and an inapproximability result of 1.5 [19].

As for $T_{\texttt{IP+EF}}$, we provided an $O(\log m)$ approximation algorithm and we know the problem is NP-hard because integral machine scheduling with identical machines is known to be NP-hard (Garey & Johnson, reduction to partition) and any assignment on identical machines is trivially locally efficient and hence EF. This leaves a wide gap as for the (in)approximability of $T_{\texttt{IP+EF}}$. Closing this gap seems challenging:

- A 2-approximation algorithm for $T_{\texttt{IP}}(c_{ij})$ was constructed using the relation to $T_{\texttt{LP}}(c_{ij})$ [19]. This approximation algorithm is based on taking a fractional schedule $a$ and rounding it to an integral one with a makespan larger by at most an additive term of $\max_{i,j|a_{ij}>0} c_{ij}$ over that of $a$, where $c_{ij}$ is the time required by machine $i$ to run job $j$. This approach does not immediately carry over, when starting from a fractional envy-free schedule, because the EF constraints might be violated when rounding.

- The inapproximability result of 1.5 for $T_{\texttt{IP}}(c_{ij})$ [19] was for makespan minimization. However, the instance used is in fact envy-free. Thus, [19] further implies that one cannot approximate the *minimal makespan and envy-free assignment* to within a factor of 1.5 in polynomial time.

- Lastly, our lower bound on the ratio $T_{\texttt{IP+EF}}(c_{ij})/T_{\texttt{IP}}(c_{ij})$ precludes obtaining a tighter approximation using a better rearrangement of the bundles of a solution to $T_{\texttt{IP}}(M)$ to achieve envy-freeness.